\documentclass[11pt]{article}
\usepackage{Blois,epsfig}

\def\be{\begin{equation}}
\def\ee{\end{equation}}
\def\bea{\begin{eqnarray}}
\def\eea{\end{eqnarray}}

\begin{document}
\vspace*{2cm}
\begin{center}
\Large{\textbf{XIth International Conference on\\ Elastic and Diffractive Scattering\\ Ch\^{a}teau de Blois, France, May 15 - 20, 2005}}
\end{center}

\vspace*{2cm}
\title{THE ALICE DETECTOR AT LHC}

\author{ R. SCHICKER }

\address{Phys. Inst., University Heidelberg, Philosophenweg 12,\\
69120 Heidelberg, Germany}

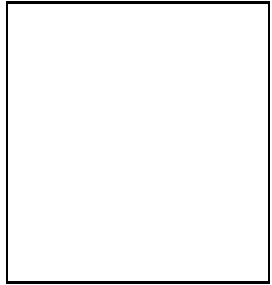
\maketitle\abstracts{
The ALICE experiment at the Large Hadron Collider LHC is presented,
and an overview of its physics program is given. A few specific 
observables are discussed in order to illustrate the physics 
potential of ALICE.
}

\section{Introduction} \label{sec:Intro}

The ALICE experiment is a general purpose heavy ion experiment
to study the behaviour of strongly interacting matter at LHC
energies. The behaviour of such matter addresses both equilibrium
and non-equilibrium physics in an energy density region of 
$\varepsilon \sim$ 1-1000 GeV/fm$^3$. This range
includes the energy densities at which phase transitions 
of strongly interacting matter occur representing sudden changes 
of the nonperturbative vacuum structure.
Of particular interest are the deconfinement transition and the 
restoration of chiral symmetry. Even though the two transitions 
are related to different aspects of the QCD vacuum, both are 
closely connected at the critical temperature $T_{c}$. 
A plethora of experimental observables is expected to indicate 
the onset of these phase transitions. It is the goal 
of the ALICE experiment to establish signatures of these 
transitions. 

Nucleus-nucleus collisions at LHC open up a new domain.
The available nucleon-nucleon center of mass energy as compared 
to RHIC is larger by a factor of about 30.
ALICE will probe parton distributions at values of 
Bjorken-x as low as 10$^{-5}$. At these values, the gluon density 
is thought to be close to phase space saturation, and strong 
nuclear gluon shadowing is expected.    

\section{The ALICE detector}

ALICE is designed as a general purpose experiment with  
a central barrel and a forward muon spectrometer \cite{Alice_prop}.
The central barrel pseudorapidity acceptance is $|\eta| \le$ 0.9
with a magnetic field of 0.5 T.
The central detectors track and identify particles 
from $\sim$ 100 MeVc$^{-1}$  to  
\mbox{$\sim$ 100 GeVc$^{-1}$} transverse momenta. Short-lived 
particles such as hyperons, D and B mesons are identified by their 
reconstructed secondary decay vertex. The detector 
granularity is chosen such that these tasks can be performed in 
a high multiplicity environment of up to 8000 charged particles 
per unit of rapidity. Tracking of particles is achieved by the 
inner tracking system (ITS) of six layers of silicon detectors, 
a large Time-Projection-Chamber (TPC) and a high granularity 
Transition-Radiation Detector (TRD). Particle identification in 
the central barrel is performed by measuring energy loss  in the 
tracking detectors, transition  radiation in the TRD and  
time-of-flight in a high-resolution TOF array. 
A single arm High-Momentum Particle Identification Detector
(HMPID) with limited solid angle coverage extends the momentum 
range of identified hadrons.
Photons will be measured 
by a crystal PbWO$_{4}$ PHOton Spectrometer (PHOS).
Additional detectors close to the beam pipe 
define an interaction trigger.
The forward muon spectrometer covers the pseudorapidity 
region -4.0 $\le \eta \le$ -2.5 with a low momentum cutoff
of 4 GeVc$^{-1}$.
 
\section{LHC experimental conditions} \label{sec:LHC}

The Large Hadron Collider LHC is designed to run in proton-proton,
proton-nucleus and nucleus-nucleus mode. In nucleus-nucleus mode, 
the Pb-Pb system is considered to be most important due to the 
highest energy densities reached in these collisions. 
Lower energy densities are reached in intermediate mass systems.
Taking data in proton-proton collisions has a threefold purpose.
First, proton-proton  collisions at LHC explore a new energy domain
and contain interesting physics in its own right \cite{Alice_phys}. 
Second, detector calibration is much simpler in the low multiplicity 
environment of proton collisions. Third, these  data serve as reference 
for nucleus-nucleus collisions  to establish medium modifications of 
observed quantities. 

\begin{table}[h]
\caption{Maximum nucleon-nucleon center of mass energy, geometric cross 
section, initial and time averaged luminosity for 
different collision systems \label{tab:exp}}
\vspace{0.4cm}
\begin{center}
\begin{tabular}{|c|c|c|c|c|}
\hline
System & $\sqrt{s}$ (TeV)  & $\sigma_{geom}$(b) & $\mathcal{L}_{0}$ 
(cm$^{-2}$s$^{-1}$) & $\mathcal{L}/\mathcal{L}_{0}$ \\ \hline
Pb-Pb & 5.5 & 7.7 & 1.0 x 10$^{27}$ &  0.44 \\ \hline
Ar-Ar & 6.3 & 2.7 & 1.0 x 10$^{29}$ &  0.64 \\ \hline
O-O & 7.0 & 1.4 & 2.0 x 10$^{29}$ &  0.73 \\ \hline
p-p & 14.0 & 0.07 & 5.0 x 10$^{30}$ &   \\ \hline
\end{tabular}
\end{center}
\end{table}

Table \ref{tab:exp} lists center of mass energies and expected 
luminosities for some systems which will be measured by ALICE. The 
time averaged luminosity  $\mathcal{L}$ depends on machine filling 
time, experiment setup time, beam lifetime and initial beam 
luminosity $\mathcal{L}_{0}$. The electromagnetic cross section  of 
$\sigma \sim$ 500 b limits the lifetime of the Pb beams to about 7 or 4 
hours for one or two data taking experiments, respectively. 
First pp collisions will be measured during the commissioning 
of the LHC. Pb-Pb collisions are expected at the end of the first pp run.

\section{The ALICE physics program}

The ALICE experimental program spans a wide range of physics topics pertinent 
to the understanding of strongly interacting matter \cite{Alice_phys}. 
Due to limitation in scope of this contribution, I have chosen three 
different topics in order  to illustrate the ALICE physics potential.

\subsection{Multiplicity distribution}

The average charged particle multiplicity per unit of rapidity 
dN$_{\rm{Ch}}$/dy is one of the first observables which will be measured 
at LHC start-up. Since the multiplicity is related to the 
entropy density and hence to the energy density,
it affects the calculation of most other observables. Despite the fundamental 
importance, there is so far no ab initio calculation of particle multiplicity 
starting from the QCD Lagrangian. The particle multiplicity  is  driven 
by soft non-perturbative QCD, and the relevant processes must be modelled by 
the new scale R$_{A} \sim$ A$^{1/3}$ fm. 

The inclusive hadron rapidity density in pp $\rightarrow$ hX is defined to be 
\begin{equation}
\frac{dN_{Ch}}{dy}= \frac{1}{\sigma_{in}^{pp}(s)} 
\int_{0}^{p_{t}^{max}} 
dp^{2}_{t}\frac{d\sigma^{pp \rightarrow hX}}{dy dp^{2}_{t}},
\label{eq:ppmult}
\end{equation}

The pp inelastic cross section $\sigma_{in}^{pp}$ grows at energies 
$\sqrt{s} > $ 20 GeV. The energy dependence is, however, poorly known, and
can be parametrized by either logarithmic or power law behaviour. 
Extrapolating measured  pp data by such fits results in an extrapolated
rapidity density in pp collisions of about 5 at energies 
relevant for ALICE \cite{pp_mult}.

The multiplicity in central nucleus-nucleus collisions at LHC energies can be 
estimated by dimensional arguments. A saturation scale Q$_{s}$ is 
assumed which represents the transverse density of all particles
produced within one unit of rapidity

\begin{equation}  
\frac{N}{R_{A}^{2}} = Q_{s}^{2}  
\label{eq:aamult1}  
\end{equation}     

where R$_{A}$ = A$^{1/3}$ fm and proportionality factors are set to unity.

In central nucleus-nucleus collisions, at scale $Q_{s}$, we have 

\begin{equation}  
N = \frac{A^{2}}{R_{A}^{2}} \frac{1}{Q_{s}^{2}}  
\label{eq:aamult2}  
\end{equation}     

with the factors $A^{2}/R_{A}^{2}$ stemming from the nuclear overlap
function and $1/Q_{s}^{2}$ from the subprocess cross section.

Eqs. \ref{eq:aamult1}  and \ref{eq:aamult2} can be combined  
with R$_{A}$= A$^{1/3}$ fm 

\begin{equation}  
N = A = Q_{s}^{2} R_{A}^{2} \rightarrow Q_{s} = 0.2 A^{1/6} \rm{GeV}.  
\label{eq:aamult3}  
\end{equation}     
 
A more refined analysis has to include the energy dependence of  
Q$_{s}$, for example by a power law dependence.
Models of LHC particle multiplicity need to determine the constant 
factors which have been set to unity in this dimensional argument.

\subsection{Parton energy loss}

After an energetic parton is produced in a medium by a hard collision, 
it will radiate energy by emitting a gluon \cite{Baier}. Both the parton 
and the gluon traverse the medium of size L. The average energy loss of the 
parton in the limit $E_{parton} \rightarrow \infty$ due to gluon radiation 
with a spectrum $\frac{\omega dI}{d\omega}$ is given by 

\begin{equation}  
\Delta E  = \int^{\omega_c}\frac{\omega dI}{d\omega} 
\simeq \alpha_{s}\omega_{c},
\hspace{2.cm} \rm{with} \hspace{1.cm} \omega_{c} = \frac{1}{2}\hat{q} L^{2}
\label{eq:eloss}  
\end{equation}     

The medium dependence of the energy loss is governed by the transport 
coefficient 

\begin{equation}  
\hat{q} \simeq \mu^{2}/\lambda \simeq \rho \int dq^{2}_{\perp} q^{2}_{\perp}
d\sigma/dq^{2}_{\perp}
\label{eq:transcoeff}  
\end{equation}     

with $\rho$ the density of the medium and $\sigma$ the gluon-medium 
interaction cross section. 

The coefficient $\hat{q}$ can be expresssed by the gluon structure 
function, i.e. for nuclear matter 

\begin{equation}  
\hat{q} = \frac{4\pi^{2}\alpha_{s} N_{c}}{N^{2}_{c}-1}\rho [xG(x,\hat{q}L)]
\label{eq:transcoeff1}  
\end{equation}     
 
with $xG(x,Q^2)$ the gluon distribution for a nucleon and $\rho$ 
the nuclear density. 

The energy loss of heavy quarks is expected to be reduced due to 
suppression of gluon radiation in a cone which scales with quark mass. 
The yield of inclusive large $p_{\perp}$ hadrons in nucleus-nucleus 
collisions is modified due to the medium induced radiative energy loss.
This induced energy loss leads to jet quenching. The quenching of jets 
can be measured by comparing spectra of particles produced in 
nucleus-nucleus and nucleon-nucleon collisions \cite{dEnterria}.
Moreover, particle spectra of nucleus-nucleus collisions 
can be analyzed with respect to the reaction plane.

\subsection{Quarkonia production}

The dissociation of quarkonia states is one of the most important 
observable for the existence of a deconfined state. The suppression 
of quarkonia is due to the shielding of potential by Debye 
screening \cite{Satz}. A quantitative characterization of dissociation 
temperature depends on the structure of heavy quark potential
which can be extracted by
an analysis of heavy quark free energies on the lattice \cite{Karsch}. 
While some authors claim an abrupt $J/\Psi$ dissociation at $T = 1.9\,T_{c}$, 
others conclude a rather gradual $J/\Psi$ dissociation with complete 
disappearance only at $T = 3.0\, T_{c}$ \cite{Asakawa,Datta}.

In order to be reliable probes in nucleus-nucleus collisions, the cross 
sections for heavy quark and quarkonia production need to be known in 
proton-proton collisions. Predictions for the 
$c\overline{c}$-pair production cross section in pp collisions at 14 TeV 
range from 7 to 17 mb depending on the values used for the charm mass and 
for the factorization and renormalization scale \cite{HQ}. 
In the case of bottom pairs in 
pp-collisions at 14 TeV, the different parameter sets result in 
$b\overline{b}$ production cross sections between 0.2 and 0.7 mb. 

Nuclear absorption and secondary scatterings with comovers can break up the 
quarkonia states and hence reduce the expected rates. The large number of 
$q\overline{q}$ pairs produced at LHC energies could, however, be an abundant 
source of final state quarkonia by coalescence due to statistical
hadronization. This mechanism  could result 
in an enhanced J/$\Psi$ production at LHC energies \cite{PBM}.

\section*{Acknowledgments}
{\small This work was supported in part by the German BMBF  under project
06HD160I. }

\end{document}